\documentclass[12pt]{article}
\usepackage{pic04}
\usepackage{hyperref}
\usepackage{url}
\usepackage{graphicx}

\begin{document}

\title{\bf POSITRON/PROTON SEPARATION USING THE AMS-02 TRD}
\author{
Gianpaolo Carosi \\
{\em MIT Lab for Nuclear Science, Bldg 44, 51 Vassar St, Cambridge, MA, USA, 02139} \\
}
\maketitle

\baselineskip=14.5pt
\begin{abstract}
AMS-02 is a cosmic ray experiment that will be placed on the International Space 
Station. One of its goals is to search for WIMP Dark Matter, specifically from 
anomalous features in the positron spectrum. In order to identify positrons at 
high energy from the large background of protons, a Transition Radiation Detector 
(TRD) will be used. Here we will present studies of positron/proton separation using the 
TRD from the AMS-02 Simulation.
\end{abstract}

\baselineskip=17pt

\section{DARK MATTER SEARCH}
One of the leading candidates for dark matter is WIMPs (Weakly Interacting 
Massive Particles) of which a favored type is the supersymmetric (SUSY) neutralino, a linear
superposition of the SUSY partners to the photon, $Z^0$, and Higgs bosons.
There is a finite cross-section for neutralinos to annihilate with each other and 
thus produce standard particles such as positrons, electrons, anti-protons, etc.
The output of such annhilations in the Milky Way halo may be detectable as a bump in 
the power-law spectrum of positrons \cite{Jung95}. 

\section{AMS TRD Simulation}

Transition radiation occurs when a highly relativistic ($\gamma>300$) charged particle 
passes through a material with varying index of refraction and emits an X-ray. A 
simulation code primarily based on Geant 3.21 is used to study the ability of the 
TRD to separate highly relativistic positrons from slower protons with the same energy 
\cite{Chou00}. The TRD geometry includes 20 
layers of radiator and straw tubes with a mixture of gaseous Xe:CO$_2$ in a ratio of 80:20 by volume.
The code simulates TR photon generation/absorption as well as 
standard $\frac{dE}{dX}$ loss in the thin Xe gas layers of the straw tubes and has 
been shown to reproduce test beam results.\cite{Sied02}

\section{Positron/Proton Separation Algorithm}

These studies used monoenergetic (50 GeV) protons/positrons generated at 
random angles and positions above the AMS detector. 
A log likelihood method was used to separate the positrons from the protons.\cite{Noma97} 
The log likelihood is defined as:
\begin{equation}\label{eq:logl}
\mathcal{L} = \sum_{i=1}^{N}log\frac{P\left(dE_i|e\right)}{P\left(dE_i|p\right)+P\left(dE_i|e\right)}
\end{equation}
where $N$ is the number of straw tube hits in a particular
event. $P\left(dE_i|e\right)$ and $P\left(dE_i|p\right)$ are
probability density functions for a positron ($e$) and proton ($p$)
to deposit energy $dE_i$ in the $i$th straw tube, respectively.
The current method of evaluating $e/p$ separation starts with creating distributions of 
energy deposited in each straw tube for protons and positrons using the AMS simulation. 
One hit is the total energy deposited in a particular straw tube for an event,
and may include $\frac{dE}{dX}$ and TR X-rays. Normalizing these histograms by the total 
number of proton (positron) hits we get the probabilities that a proton (positron) 
will deposit a specific amount of energy in a straw tube. Next we use the log likelihood 
ratio estimator in Equation 1 to create distributions of $\mathcal{L}$, for
both positron and proton events.

\begin{figure}[htbp]
  \centerline{\hbox{ \hspace{0.2cm}
    \includegraphics[width=6.5cm,height=3.4cm]{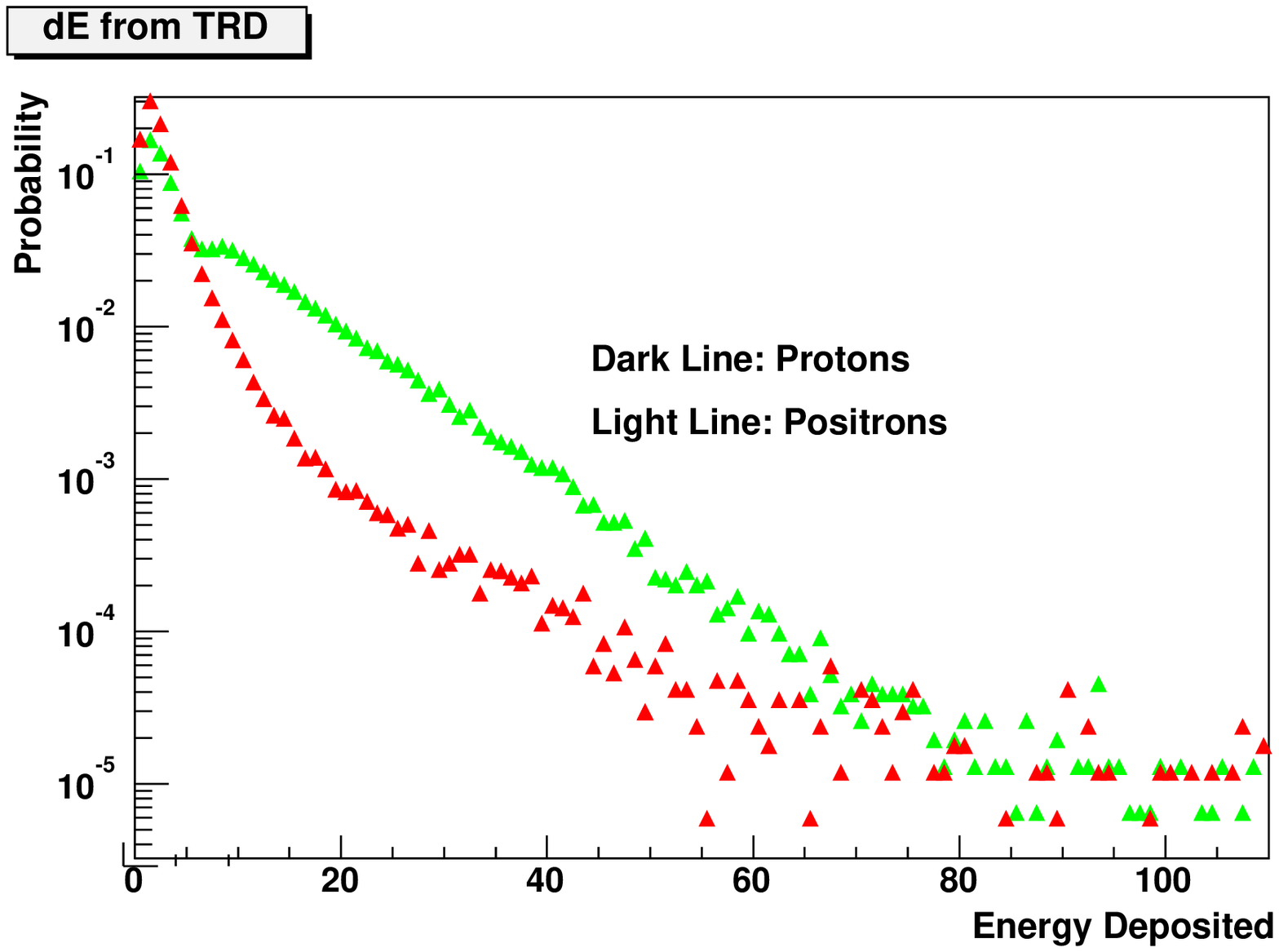}
    \hspace{0.3cm}
    \includegraphics[width=6.5cm,height=3.4cm]{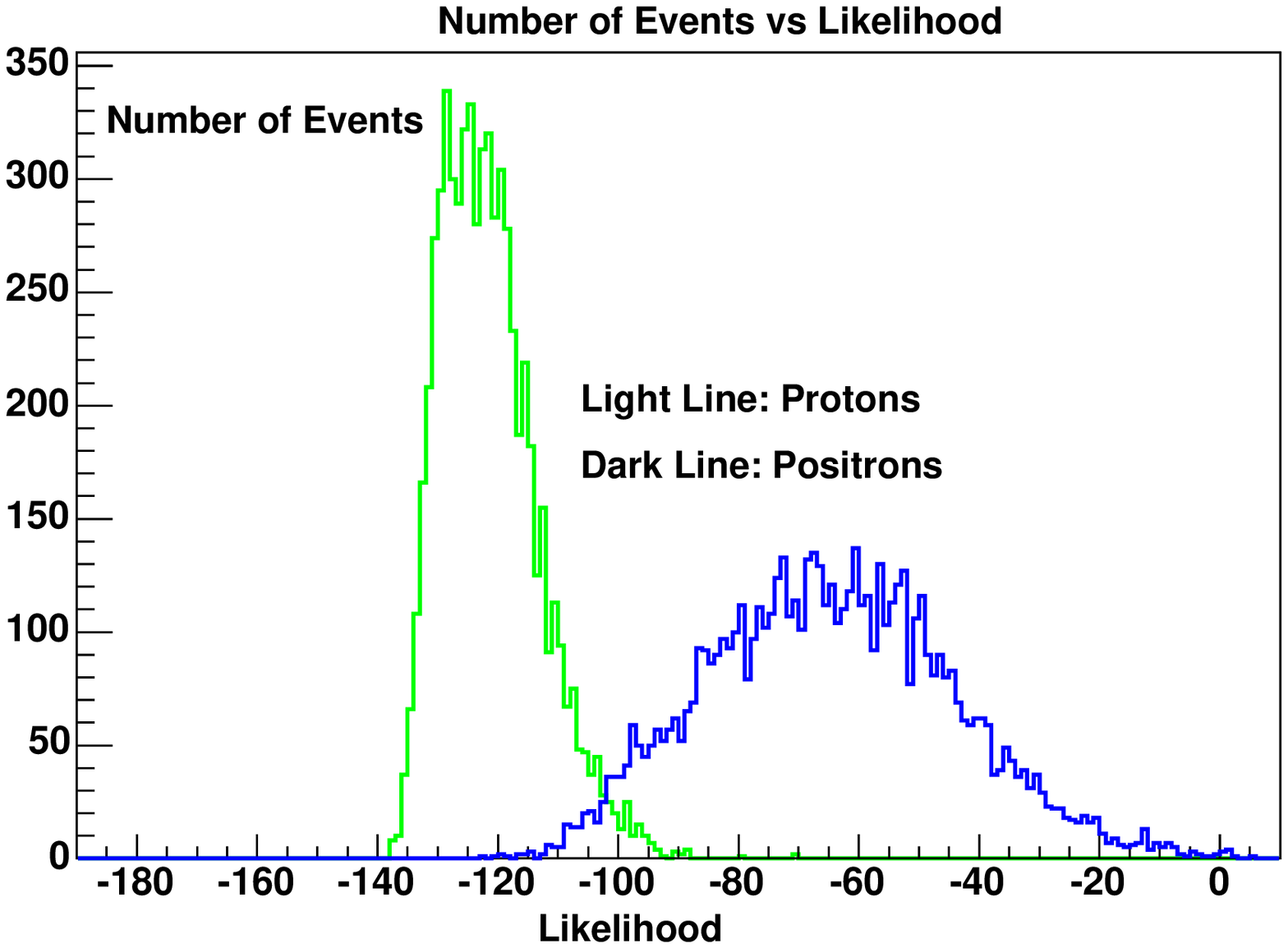}
    }
  }
 \caption{\it
   Probability of Energy Deposited and Likelihood Ratio Estimator
    \label{twofig} }
\end{figure}

Then the fraction of positrons (protons) above a threshold likelihood 
$\mathcal{L}_{th}$ defines the positron efficiency (proton
contamination). Integrating to the right of $\mathcal{L}_{th}$
determines the efficiency vs $\mathcal{L}_{th}$. Our goal
is to minimize the proton contamination while keeping a reasonable
positron efficiency. Finally we plot proton contamination vs positron 
efficiency, where each point of the plot corresponds to a
different $\mathcal{L}_{th}$. To compare particle separation qualities 
for different configurations, we choose
a threshold likelihood for which 90\% of the positrons satisfy the
likelihood cut.

\begin{figure}[htbp]
  \centerline{\hbox{ \hspace{0.2cm}
    \includegraphics[width=6.5cm,height=3.4cm]{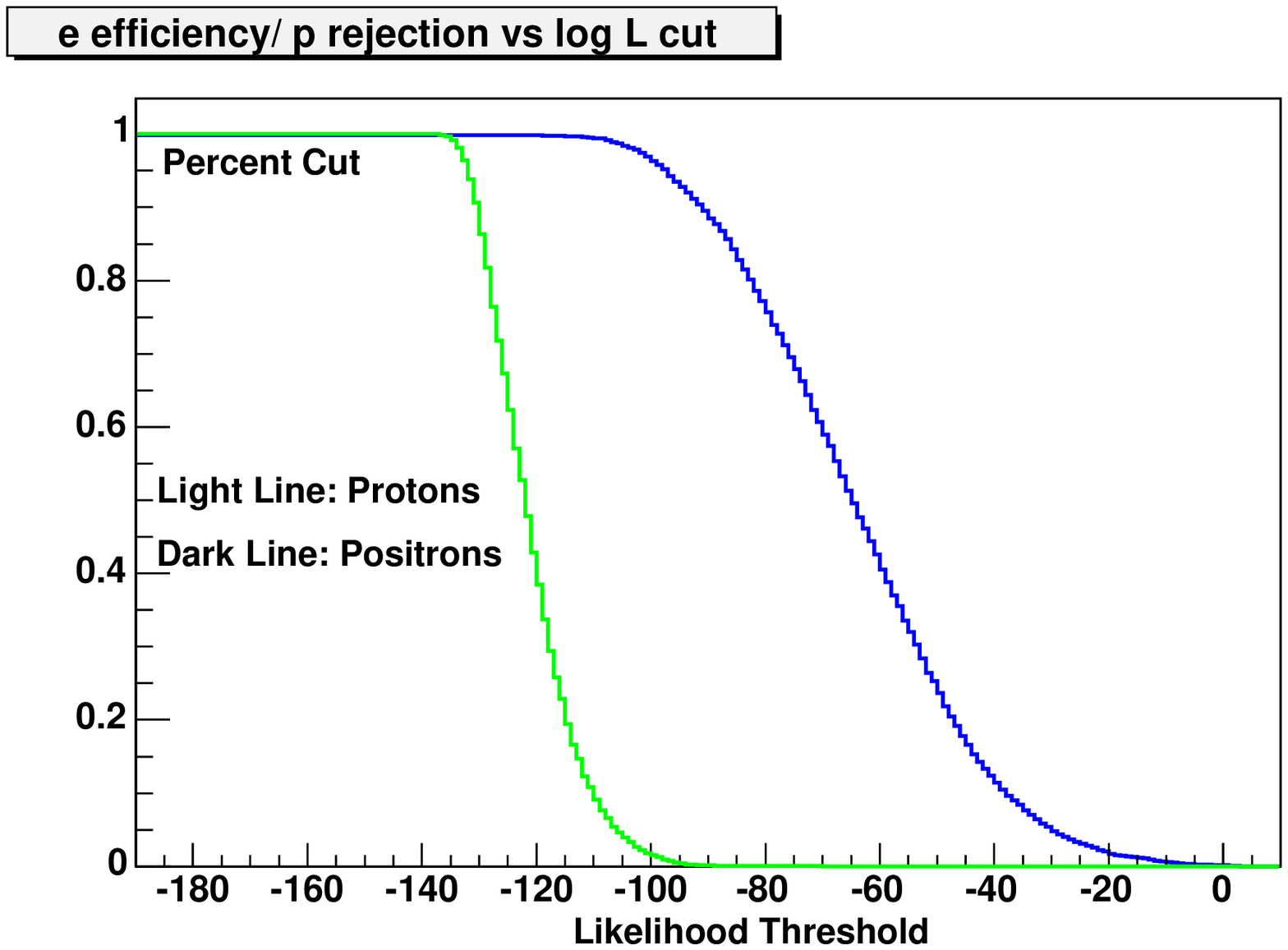}
    \hspace{0.3cm}
    \includegraphics[width=6.5cm,height=3.4cm]{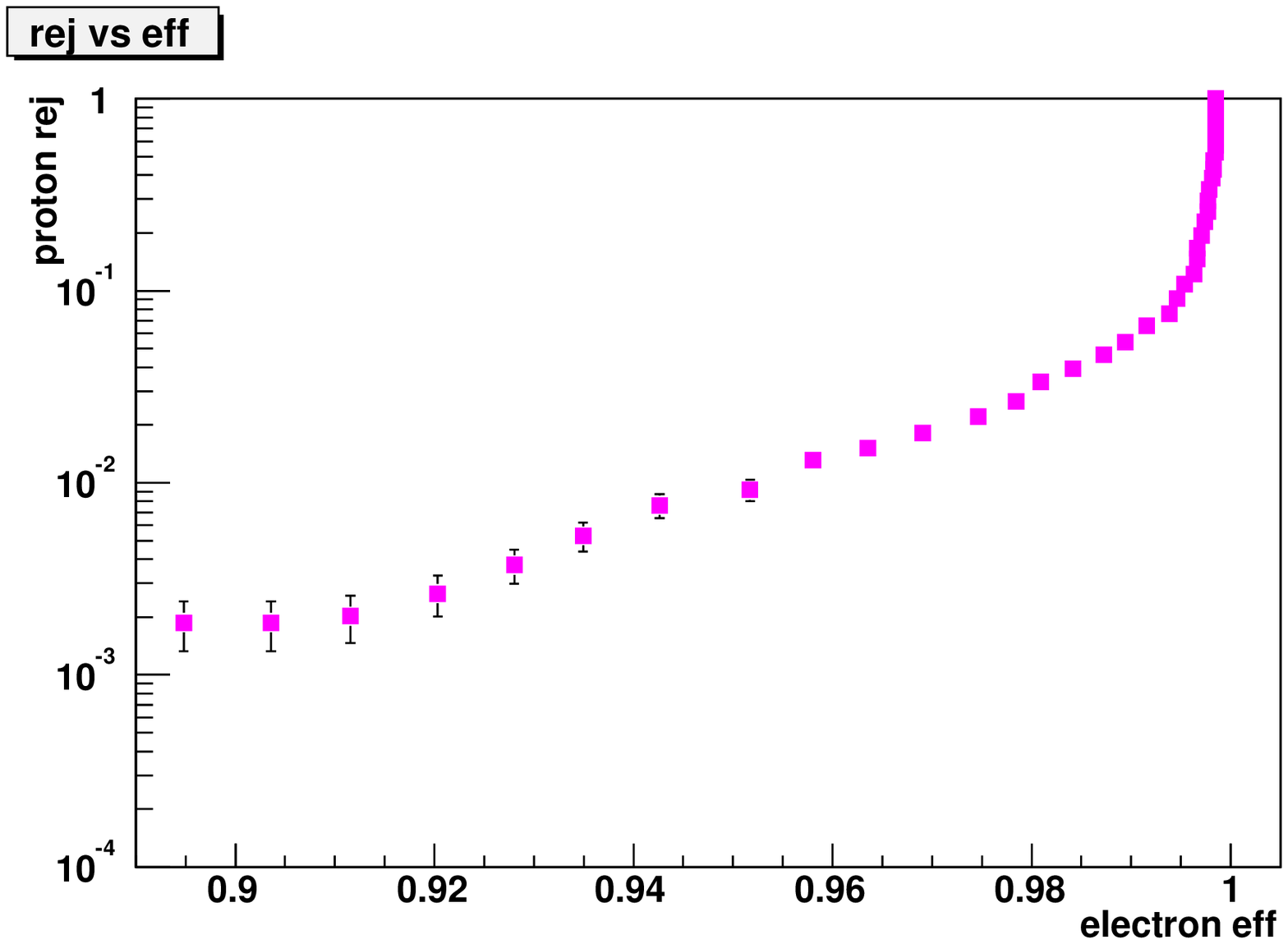}
    }
  }
 \caption{\it
   Integrated Efficiency and Proton Contamination vs Positron Efficiency
    \label{twofig} }
\end{figure}


\section{STUDIES AND RESULTS}

Initial results showed that proton rejections of approximately
$10^{-3}$ could be achieved while only throwing away 10\% of the positron signal. 
This was obtained by using all the hits in each event. Studies were conducted 
at different particle momentum. The results agreed with the expectation that 
the separation became worse as the momentum rose and protons themselves began to 
emit transition radiation. Removing even 2 layers had a drastically negative 
affect on the separation (factor of 1.5 at 100 GeV). Studies were also 
conducted to determine how noise in the gas-gain would 
affect the positron/proton separation. Uncorrelated 
variations in the gas-gain had a small effect on separation until they
reached 30\% of the total gain.

\end{document}